\documentclass[a4paper]{jpconf}
\usepackage{graphicx}
\usepackage{amsmath,amssymb,amsfonts,amsthm}
\def\lsim{\hbox{ \raise.35ex\rlap{$<$}\lower.6ex\hbox{$\sim$}\ }}
\def\gsim{\hbox{ \raise.35ex\rlap{$>$}\lower.6ex\hbox{$\sim$}\ }} 

\begin{document}
\title{Phenomenology of loop quantum cosmology}

\author{Mairi Sakellariadou}
    
\address{Department of Physics, King's College London, University of
  London, Strand WC2R 2LS, UK}

\ead{Mairi.Sakellariadou@kcl.ac.uk}

\begin{abstract}
After introducing the basic ingredients of Loop Quantum Cosmology, I
will briefly discuss some of its phenomenological aspects. Those can
give some useful insight about the full Loop Quantum Gravity theory
and provide an answer to some long-standing questions in early
universe cosmology.
\end{abstract}

\section{Introduction}
The variety of precise astrophysical and cosmological data, available
at present, combined with the large number of high energy physics
experiments, are expected to give us the necessary ingredients to test
fundamental theories and understand the very early phases of the
evolution of our universe.

Cosmological inflation~\cite{infl} remains the most promising
candidate to solve the shortcomings of the standard hot big bang
model, while it offers a causal explanation for the primordial
fluctuations with the correct Cosmic Microwave Background (CMB)
features.  Despite this success, inflation suffers from a number of
drawbacks~\cite{drawbacks}.  In particular, compatibility between
theory and measurements often necessitates fine-tuning of the
inflationary parameters, inflation remains still a paradigm in search
of a model, while it must prove itself
generic~\cite{generic1,generic2,generic3,generic4,generic5,generic6,generic7}.

There is undoubtfully an additional list of fundamental cosmological
questions, still lacking a satisfying answer. One does not know, for
instance, how close to the big bang a smooth space-time can be
considered as the correct framework. Quantum gravity, a full theory
which is supposed to resolve the big bang singularity, is still
missing, while it is not known whether a new boundary condition is
needed at the big bang, or whether quantum dynamical equations remain
well-behaved at singularities.  It is nevertheless clear that a smooth
space-time background cannot be assumed as the correct description
close to the big bang.

In a Hamiltonian formulation to a quantum theory, the absence of
background metric indicates that Hamiltonian dynamics is generated by
constraints. Physical states are solutions to quantum constraints
implying that all physical laws are obtained from these solutions,
while there is no external time according which evolution can be
studied.  One has to define a monotonic variable to play the r\^ole of
emergent, or internal, time, and then build a framework within which
the short-distance drawbacks of General Relativity near the big bang
can be cured, maintaining however the agreement with General
Relativity at large scales.

Loop Quantum Gravity (LQG)~\cite{rovelli04,thiemann07} is a
non-perturbative and background independent canonical quantisation of
General Relativity in four space-time dimensions. Loop Quantum
Cosmology (LQC)~\cite{bojowald05} is a cosmological mini-superspace
model quantised with methods of LQG.  The discreteness of spatial
geometry and the simplicity of the setting allow for a complete study
of dynamics.  The difference between LQC and other approaches of
quantum cosmology, is that the input is motivated by a full quantum
gravity theory. The simplicity of the setting, combined with the
discreteness of spatial geometry provided by LQG, render feasible the
overall study of LQC dynamics. 

In what follows, I will briefly discuss some of the phenomenological
consequences of LQC, a subject which gains constantly an increasing
interest from the scientific community, due in particular to its
successes.

\section{Elements of LQG/LQC}

Loop quantum cosmology is formulated in terms of SU(2) holonomies of
the connection and triads.  The canonically conjugated variables are
defined by the densitised triad $E^a_i$, and an SU(2) valued
connection $A^i_a$, where $i$ refers to the Lie algebra index, and $a$
is a spatial index with $a, i= 1, 2, 3$. The densitised triad gives
information about spatial geometry ({\sl i.e.}, the three metric),
while the connection gives information about curvature (spatial and
extrinsic one).  Let me explain this: Quantum gravity introduces a
discreteness to space-time. To quantise quantum gravity, this
discreteness manifests itself as quanta of space.  Since the quanta
are three-dimensional, they can be characterised by a triad of
numbers. The connection between adjacent pairs of these {\sl quanta}
form a two-dimensional surface, through which the geometric connection
is defined.  To ensure that local rotations do not induce different
geometries, this connection must be SO(3) invariant.  By using
connection-triad variables, arising from a canonical transformation of
Arnowitt-Desner-Misner ( variables, one can make an analogy with
gauge theories, which is particularly useful when dealing with
quantisation issues.

For any quantisation scheme based on a Hamiltonian framework ({\sl
  e.g.}, LQC) or an action principle ({\sl e.g.}, path integral) for a
homogeneous and flat model, divergences appear which need to be
regularised. To remove the divergences that occur in non-compact
topologies, the spatial homogeneity and Hamiltonian are restricted to
a finite fiducial cell, of scale $\mu_0$, with finite volume
$V_0=\int{\rm d}^3x\sqrt{|^0q|}$, where $^0q$ is the determinant of
the fiducial background metric.

In the case of a spatially flat background, derived from the Bianchi
I model, the isotropic connection can be expressed in terms of the
dynamical component of the connection ${\tilde c}(t)$ as 
\begin{equation}
A_a^i={\tilde c}(t)\omega_a^i~, 
\end{equation}
with $\omega_a^i$ a basis of left-invariant one-forms $\omega_a^i={\rm
  d}x^i$.  The densitised triad can be decomposed using the Bianchi I
basis vector fields $X_i^a=\delta_i^a$ as 
\begin{equation}
E_i^a=\sqrt{^0q}{\tilde p}(t)X_i^a~, 
\end{equation}
where $^0q$ stands for the determinant of the fiducial background
metric $^0q_{ab}=\omega^i_a\omega_{bi}$, and ${\tilde p}(t)$ denotes
the remaining dynamical quantity after symmetry reduction.

In terms of the metric variables with three-metric
$q_{ab}=a^2\omega_a^i\omega_{bi}$, the dynamical quantity is just the
scale factor $a(t)$. Given that the Bianchi I basis vectors are
$X_i^a=\delta^a_i$, 
\begin{equation}
|\tilde{p}|=a^2~, 
\end{equation}
where the absolute value is taken because the triad has an
orientation.  Since the basis vector fields are spatially constant in
the spatially flat model, the connection component is 
\begin{equation} 
{\tilde c}={\rm sgn}({\tilde p})\gamma \frac{\dot a}{N}~;
\end{equation} 
$N$ is the lapse function and $\gamma$ the Barbero-Immirzi parameter,
a quantum ambiguity parameter approximately equal to 0.2375.

The canonical variables ${\tilde c}, {\tilde p}$ are related through
the Poisson bracket 
\begin{equation}
\{{\tilde c}, {\tilde p}\}=\frac{\kappa\gamma}{3}V_0~,
\end{equation} 
where $V_0$ the volume of the elementary cell adapted to the fiducial
triad and $\kappa\equiv 8\pi G$.

Defining the triad component $p$, determining the physical volume of
the fiducial cell, and the connection component $c$, determining the
rate of change of the physical edge length of the fiducial cell, as
\begin{equation}
p=V_0^{2/3}{\tilde p}~~~~~,~~~~~c=V_0^{1/3}{\tilde c}~,
\end{equation}
respectively, we obtain
\begin{equation}
\{c,p\}=\frac{\kappa\gamma}{3}~,
\end{equation}
independent of the volume $V_0$ of the fiducial cell.

To quantise the theory one faces the usual difficulty of quantum
cosmology. Namely, the metric itself has to be considered as a
physical field, thus it must be quantised; it is not a fixed
background.  Any {\sl standard} quantisation scheme fails: Fock
quantisation fails, while even free Hamiltonians are non-quadratic in
metric dependence, so there is no simple perturbation analysis. The
way out is to use gauge theory variables to define holonomies of the
connection along a given edge
\begin{equation}
h_e(A)={\cal P}\exp\int{\rm d}s\dot\gamma^\mu(s)A_\mu^i(\gamma(s))\tau_i~,
\end{equation}
where ${\cal P}$ indicates a path ordering of the exponential,
$\gamma^\mu$ is a vector tangent to the edge and
$\tau_i=-i\sigma_i/2$, with $\sigma_i$ the Pauli spin matrices, and
fluxes of a triad along an $S$ surface
\begin{equation}
E(S,f)=\int_S\epsilon_{abc}E^{ci}f_i{\rm d}x^a{\rm d}x^b~,
\end{equation}
with $f_i$ an SU(2) valued test function.  Certainly, at the level of
LQC these variables may seem ad-hoc and unnatural, however their
motivation follows naturally from the full LQG theory.

Thus, the basic configuration variables in LQC are holonomies of the
connection 
\begin{equation}
h_i^{(\mu_0)}(A)=\cos\left({\mu_0c\over 2}\right){\bf 1}+2\sin\left({\mu_0c\over
  2}\right)\tau_i ~,
\end{equation}
along a line segment $\mu_0\ ^0e^a_i$ and the flux of the triad
$$F_S(E,f)\propto p~,$$ where the basic momentum variable is the triad
component $p$, ${\bf 1}$ is the identity $2\times 2$ matrix and
$\tau_i=-i\sigma_i/2$ is a basis in the Lie algebra SU(2) satisfying
the relation
$$\tau_i\tau_j=(1/2)\epsilon_{ijk}\tau^k-(1/4)\delta_{ij}~.$$

In the classical theory, curvature can be expressed as a limit of the
holonomies around a loop as the area enclosed by the loop shrinks to
zero. In quantum geometry however, the loop cannot be continuously
shrunk to zero area and the eigenvalues of the area operator are
discrete. Thus, ∆there is a smallest non-zero eigenvalue, the area gap
$\Delta$~\cite{al}. As a result, the Wheeler-de Witt 
differential equation gets replaced by a difference equation whose
step size is controlled by $\Delta$.

Let me start with the {\sl old} quantisation procedure. Along the
lines of LQG, one considers $e^{i\mu_0c/2}$ (with $\mu_0$ an arbitrary
real number) and $p$, as the elementary classical variables, which
have well-defined analogues.  Using the Dirac bra-ket
notation and setting $e^{i\mu_0c/2}=\langle c|\mu\rangle$, the action
of the operator $\hat p$ acting on the basis states $|\mu\rangle$
reads
\begin{equation} 
\hat p|\mu\rangle=\frac{\kappa\gamma\hbar|\mu|}{6}|\mu\rangle~, 
\end{equation}
where $\mu$ (a real number) stands for the eigenstates of $\hat p$,
satisfying the orthonormality relation 
$\langle\mu_1|\mu_2\rangle=\delta_{\mu_1,\mu_2}$.  
The action of the $\widehat{\exp \left[ \frac{i\mu_0}{2} c\right]} $
operator acting on basis states $|\mu\rangle$ is 
\begin{equation} 
\widehat{\exp \left[ \frac{i\mu_0}{2} c \right]} | \mu \rangle = \exp
\left[ \mu_0 \frac{{\rm d}}{{\rm d} \mu} \right] | \mu \rangle = |
\mu+\mu_0 \rangle~, 
\end{equation} 
where $\mu_0$ is any real number.  Thus, in the {\sl old} quantisation, the
operator $e^{i\mu_0c/2}$ acts as a simple shift operator.  As in the full
LQG theory, there is no operator corresponding to the connection, however the
action of its holonomy is well defined.  The action of the
holonomies, $\hat{h}_i^{(\mu_0)}$, of the gravitational connection on
the basis states is given by~\cite{Ashtekar:2006wn}
\begin{equation}
\label{eq:hol1}
\hat{h}_i^{(\mu_0)}| \mu \rangle = \left(\widehat{\rm cs} {\bf 1} 
+ 2 \widehat{\rm sn}\tau_{\rm i} \right) | \mu
\rangle~,
\end{equation}
where,
\begin{eqnarray}
\label{defin}
 \widehat{\rm cs} |\mu\rangle \equiv \widehat{\cos (\mu_0 c/2)} | \mu
 \rangle &=& \left[ \ | \mu+\mu_0\rangle + | \mu -\mu_0\rangle
   \ \right]/2~, \nonumber \\ \widehat{\rm sn} |\mu\rangle \equiv
 \widehat{\sin (\mu_0 c/2)} | \mu \rangle &=& \left[ \ |
   \mu+\mu_0\rangle - | \mu -\mu_0\rangle \ \right]/(2i)~.  
\end{eqnarray}

The gravitational part of the Hamiltonian operator in terms of SU(2)
holonomies and the triad component, in the irreducible $J=1/2$
representation, reads~\cite{Ashtekar:2006wn}
\begin{equation}
\label{eq:ham_g1} 
\hat{\mathcal C}_{\rm g} = \frac{2i}{\kappa^2 \hbar \gamma^3
  \mu_0^3}{\rm tr} \sum_{ i j k} \epsilon^{ijk} \left(
\hat{h}_i^{(\mu_0)}\hat{h}_j^{(\mu_0)}\hat{h}i^{(\mu_0)-1}
\hat{h}_j^{(\mu_0)-1} \hat{h}_k^{(\mu_0)} \left[
  \hat{h}_k^{(\mu_0)-1},\hat{V} \right]\right){\rm sgn}(\hat p)~.
\end{equation}
The action of the self-adjoint Hamiltonian constraint operator,
$\hat{\mathcal H}_{\rm g}=(\hat{\mathcal C}_{\rm g} + \hat{\mathcal
C}_{\rm g}^{\dagger} )/2$ on the basis states $|\mu\rangle$ is
\begin{equation}
\hat{\mathcal H}_{\rm g} | \mu \rangle 
=
\frac{3}{4\kappa^2 \gamma^3 \hbar \mu_0^3}
\Bigl\{ \bigl[ R(\mu)+R(\mu+4\mu_0)\bigr]| \mu+4\mu_0 \rangle
-4R(\mu ) | \mu\rangle +\bigl[ R(\mu)+R(\mu-4\mu_0)\bigr]| 
\mu-4\mu_0 \rangle \Bigr\}~,
\end{equation}
where 
\begin{equation}
R(\mu )= \left( \kappa \gamma \hbar/6\right)^{3/2} 
\Big| | \mu+\mu_0 |^{3/2} -|\mu-\mu_0|^{3/2} \Big|~.
\end{equation}
The dynamics are then determined by the Hamiltonian constraint
\begin{equation}
(\hat{\mathcal H}_{\rm g}+\hat{\mathcal H}_\phi)|\Psi\rangle=0~,
\end{equation}
where $\hat{\mathcal H}_\phi$ stands for the matter Hamiltonian.
In the full LQG theory there is an infinite number of constraints,
while in LQC there is only one integrated Hamiltonian constraint.
Matter is introduced by just adding the actions of matter components
to the gravitational action. We can finally obtain difference
equations analogous to the differential Wheeler-de Witt equations.

More precisely, the constraint equation on the physical wave-functions
$|\Psi\rangle$, which can be expanded using the basis states as
$|\Psi\rangle = \sum_\mu \Psi_\mu(\phi)|\mu\rangle$ with summation
over values of $\mu$ and where the dependence of the coefficients on
$\phi$ represents the matter degrees of freedom,
reads~\cite{Vandersloot:2005kh}
\begin{eqnarray}\label{qee}
&&~~\Biggl[\Big| V_{\mu+5\mu_0}-V_{\mu+3\mu_0}\Big|+\Big|V_{\mu+\mu_0}
 - V_{\mu-\mu_0}\Big|\Biggr] \Psi_{\mu+4\mu_0}(\phi) -
 4\Big|V_{\mu+\mu_0}V_{\mu-\mu_0}\Big| \Psi_\mu(\phi) \nonumber \\
 &&+\Biggl[\Big|V_{\mu-3\mu_0}-
 V_{\mu-5\mu_0}\Big|+\Big|V_{\mu+\mu_0}- V_{\mu-\mu_0}\Big|\Biggr]
 \Psi_{\mu-4\mu_0}(\phi) =- \frac{4\kappa^2 \gamma^3 \hbar \mu_0^3}{3}
 {\mathcal H}_{\rm \phi}(\mu)\Psi_\mu(\phi)~;
\end{eqnarray}
the matter Hamiltonian $\hat{\mathcal H}_{\rm \phi}$ is assumed to act
diagonally on the basis states with eigenvalues ${\mathcal H}_{\rm
  \phi}(\mu)$.  Equation (\ref{qee}) is the quantum evolution (in
internal time $\mu$) equation; there is no continuous variable (the
scale factor in classical cosmology), but a label $\mu$ with discrete
steps.  The wave-function $\Psi_\mu(\phi)$, depending on internal time
$\mu$ and matter fields $\phi$, determines the dependence of matter
fields on the evolution of the universe, with a massless scalar field
playing the r\^ole of emergent time.  Thus, in LQC the quantum
evolution is governed by a second order difference equation, rather
than the second order differential equation of the Wheeler-de Witt
quantum cosmology. As the universe becomes large and enters the
semi-classical regime, its evolution can be approximated by the
differential Wheeler-de Witt equation.

\section{Lattice refinement}

The quantised holonomies were at first assumed to be shift operators
with a fixed magnitude, leading to a quantised Hamiltonian constraint
being a difference equation with a constant interval between points on
the lattice. While these models can be used to study certain aspects
of the quantum regime, they were found to lead to serious
instabilities in the continuum, semi-classical
limit~\cite{Rosen:2006bga,Bojowald:2007ra}.  Considering the continuum
limit of the Hamiltonian constraint operator, we have so far assumed
that $\Psi$ does not vary much on scales of the order of $4\mu_0$
(known as {\sl pre-classicality}), so that one can smoothly interpolate
between the points on the discrete function $\Psi_\mu(\phi)$ and
approximate them by the continuous function $\Psi(\mu,\phi)$.  Under
this assumption, the difference equation can be very well approximated
by a differential equation for a continuous wave-function.  However,
the form of the wave-function indicates that the period of
oscillations can decrease as the scale increases. Thus, the assumption
of {\sl pre-classicality} can break down at large scales, leading to
deviations from the classical behaviour. 

Let me explain this point a bit further: In the underlying full LQG
theory, the contributions to the (discrete) Hamiltonian operator
depend on the state which describes the universe. As the volume grows
({\i.e.}, the universe expands), the number of contributions
increases. Thus, the Hamiltonian constraint operator is expected to
create new vertices of a lattice state (in addition to changing their
edge labels), which in LQC results in a refinement of the discrete
lattice~\cite{ms-review}. Lattice refinement is also required from
phenomenological reasons~\cite{Bojowald:2007ra,Nelson:2007um}; for
instance, it renders a successful inflationary era more
natural~\cite{Nelson:2007um}. The effect of lattice refinement has
been modelled and the elimination of the instabilities in the
continuum era has been explicitly shown.

The appropriate lattice refinement model should be obtained from the
full LQG theory. In principle, one should use the full Hamiltonian
constraint and find the way that its action balances the creation of
new vertices as the volume increases. Instead, phenomenological
arguments have been used, where the choice of the lattice refinement
model is constrained by the form of the matter Hamiltonian~\cite{ns2}.
In particular, LQC can generically support inflation, and other matter
fields, without the onset of large scale quantum gravity corrections,
only for a particular model of lattice refinement~\cite{ns2}. This
choice is the only one for which physical quantities are independent
of the elementary cell adopted to regulate spatial
integrations~\cite{gs08}, and moreover, it is exactly the choice
required for the uniqueness of the factor ordering of the Wheeler-de
Witt equation~\cite{ns4}.

Allowing the length scale of the holonomies to vary dynamically, the
form of the difference equation, describing the evolution, changes.
Since the parameter $\mu_0$ determines the step-size of the difference
equation, assuming the lattice size is growing, the step-size of the
difference equation is not constant in the original triad variables.
Let us consider the particular case of
\begin{equation} 
\mu_0 \rightarrow\tilde{\mu}\left(\mu\right)=\mu_0\mu^{-1/2}~,
\end{equation} 
suggested by certain intuitive heuristic approaches, such as noting
that the minimum area used to regulate the holonomies should be a
physical area~\cite{Ashtekar:2006uz2}, or that the discrete step size
of the difference equation should always be of the order of the Planck
volume. Moreover, this choice also results in a significant
simplification of the difference equation, compared to more general
lattice refinement schemes.  The basic operators are given by
replacing $\mu_0$ with $\tilde{\mu}$.  Upon
quantisation~\cite{Ashtekar:2006uz2}
\begin{equation}
\widehat{ e^{i\tilde{\mu}c/2}}
|\mu \rangle=e^{-i\tilde{\mu}\frac{\rm d}{{\rm d}\mu} }|\mu\rangle~,
\end{equation}
which is no longer a simple shift operator since $\tilde{\mu}$ is a 
function of $\mu$. Changing the basis to 
\begin{equation}\label{eq:basis}
 \nu = \mu_0 \int \frac{{\rm d}\mu}{\tilde{\mu}}=\frac{2}{3} \mu^{3/2}~,
\end{equation}
one gets
\begin{equation}
 e^{-i\tilde{\mu}\frac{\rm d}{{\rm d} \mu}}|\nu\rangle 
= e^{-i\mu_0 \frac{\rm d}{
{\rm d}\nu}}|\nu\rangle = |\nu+\mu_0\rangle~.
\end{equation}
Thus, in the new variables the holonomies act as simple shift
operators, with parameter length $\mu_0$. In this sense, the basis
$|\nu\rangle$ is a much more natural choice than $|\mu\rangle$. One
can then proceed as in the previous case of a fixed spatial lattice
and write down the Hamiltonian constraint.

\subsection{Constraints on inflation}

Let me consider first a fixed and then a dynamically varying lattice
and solve the second order difference equation in the continuum
limit. Two constraints can be imposed on the inflaton potential: the
first one so that the continuum approximation is valid, and the second
one so that there is agreement with the CMB measurements on large
angular scales. A combination of the two constraints in the context of
a particular inflationary model, will give us~\cite{ns2} the
conditions for natural and successful inflation within LQC.

More precisely, let us separate the wave-function $\Psi(p,\phi)$ into
$\Psi(p,\phi)= \Upsilon(p)\Phi(\phi)$ and approximate the dynamics of
the inflaton field, $\phi$, by setting $V(\phi) =V_\phi
p^{\delta-3/2}$, where $V_\phi$ is a constant and $\delta=3/2$ in the
case of slow-roll, to get~\cite{ns2}
\begin{equation}\label{eq-p} 
p^{-1/2}\frac{{\rm d}}{{\rm d} p} \left[ p^{-1/2} \frac{{\rm d}}{{\rm
      d} p } \left( p^{3/2} \Upsilon\left(p\right) \right) \right] +
\beta V_\phi p^\delta\Upsilon(p) =0~,
\end{equation}
with solutions~\cite{ns2}
\begin{equation}
 \Upsilon\left( p\right) \approx  p^{-(9+2\delta)/8} \sqrt{\frac{
2\delta+3}{2\sqrt{\beta V_\phi}\pi} }
\Biggl[ C_1 \cos \left( x-\frac{3\pi}{2(2\delta+3)}
-\frac{\pi}{4} \right)
+C_2 \sin \left( x -\frac{3\pi}{2(2\delta+3)}-\frac{\pi}{4} \right)
\Biggr]~,
\end{equation}
where 
$x=4\sqrt{\beta V_\phi}(2\delta+3)^{-1} p^{(2\delta+3)/4}$
and $\beta=96/(\kappa\hbar^2)$.  

Without lattice refinement, the discrete nature of the underlying
lattice would eventually be unable to support the oscillations and the
assumption of {\sl pre-classicality} will break down, implying that the
discrete nature of space-time becomes significant on large scales. For
the end of inflation to be describable using classical General
Relativity, it must end before a scale, at which the assumption of
{\sl pre-classicality} breaks down and the semi-classical description is no
longer valid, is reached. Let me quantify this constraint: The
separation between two successive zeros of $ \Upsilon\left( p\right) $
is
\begin{equation}
 \Delta p =\frac{\pi}{\sqrt{\beta V_\phi}} p^{(1-2\delta)/4}~.  
\end{equation}
For the continuum limit to be valid, the wave-function must vary
slowly on scales of the order of $\mu_c = 4\tilde{\mu}$,
leading to~\cite{ns2}
\begin{equation}
\Delta p > 4\mu_0\left(\frac{\kappa\gamma\hbar}{6}\right)^{3/2}p^{-1/2}~,
\end{equation}
which implies~\cite{ns2}
\begin{equation}
 V_\phi < \frac{27\pi^2}{192\mu_0^2\gamma^3 \kappa^2\hbar} 
p^{(3-2\delta)/2}~.
\end{equation}
Assuming slow-roll inflation, we set $\delta\approx 3/2$.  Setting
$\mu_0=3\sqrt{3}/2$, the constraint on the inflationary potential, in
units of $\hbar=1$, reads
\begin{equation}
V(\phi) \lsim 2.35\times10^{-2} l_{\rm Pl}^{-4}~,
\end{equation}
which is a weaker constraint than the one imposed for fixed lattices,
namely~\cite{ns2}
\begin{equation}
V_\phi \ll 10^{-28}l_{\rm pl}^{-4}~,
\end{equation}
assuming that half of the inflationary era takes place during the
classical era.

One can further constrain the inflationary potential so that the
fractional over-density in Fourier space and at horizon crossing is
consistent with the COBE-DMR measurements, namely
\begin{equation}
\label{cobe}
 \frac{\left[V(\phi)\right]^{3/2}}{V'(\phi)}
\approx 5.2\times 10^{-4} M_{\rm Pl}^3~.
\end{equation}
To do so, one must however adopt a particular inflationary model.  As
such, let us select $V(\phi)=m^2\phi^2/2$.  Combining the two
constraints we obtain~\cite{ns2}
\begin{equation}
\label{constraint_on_m}
 m \lsim  70 (e^{-2N_{\rm cl}}) ~M_{\rm Pl}~~~~
\mbox {and}~~~~ m \lsim 10 ~M_{\rm Pl}~,
\end{equation}
for the fixed and varying lattices, respectively.

In conclusion, the requirement that a significant proportion of a
successful (in particular with respect to the CMB measurements)
inflationary regime takes place during the classical era, imposes a
strong constraint on the inflaton mass. Such a constraint turns out
however to be much softer, and therefore rather natural, once lattice
refinement is considered.  In this sense, we argue~\cite{ns2} that
lattice refinement is essential to achieve a successful inflationary
era, provided inflation can be generically ({\sl i.e.}, with generic
initial conditions) realised.

\subsection{Constraints on the matter Hamiltonian}

Given a lattice refinement model, we will show~\cite{Nelson:2007um}
that only certain types of matter can be allowed. To do so, we
parametrise the lattice refinement by $A$ and the matter Hamiltonian
by $\delta$ and solve the Hamiltonian constraint. The restrictions on
the two-dimensional parameter space will become apparent once physical
restrictions to the solutions of the wave-functions are
imposed~\cite{Nelson:2007um}.

To be more precise, consider
\begin{equation}
\tilde{\mu}=\mu_0 \mu^A~,
\end{equation}
leading to
\begin{equation}
\label{def:nu}
\nu=\frac{\tilde\mu_0\mu^{1-A}}{\mu_0(1-A)}~.
\end{equation}
Being only interested in the large scale limit, we
approximate~\cite{Nelson:2007um} the matter Hamiltonian by $\hat{\cal
  H}_\phi= \hat{\nu}^\delta\hat{\epsilon}\left(\phi\right)$, implying
\begin{equation}
\hat{\epsilon}\left(\phi\right)\Psi \equiv
\epsilon\left(\phi\right)\Psi = -\nu^{-\delta}\hat{\cal H}_{\rm g}
\Psi~.  
\end{equation}
A necessary condition so that the wave-functions are physical, is that
the finite norm of the physical wave-functions, defined by
$\int_{\phi=\phi_0}d\nu |\nu |^\delta \overline{\Psi}_1\Psi_2$, is
independent of the choice of $\phi=\phi_0$. The solutions of the
constraint are renormalisable provided they decay, on large scales,
faster than $\nu^{-1/(2\delta)}$.

To solve the constraint equation, we need to specify the from of
${\cal H}_\phi$, which has in general two terms with different scale
dependence. Being interested in the large scale limit, there is one
dominant term, allowing to write~\cite{Nelson:2007um}
\begin{equation}
\beta{\mathcal H}_{\phi} =\epsilon_\nu(\phi) \nu^{\delta_\nu}~,
\end{equation}
where the function $\epsilon_\nu$ is constant with respect to $\nu$.

Solving~\cite{Nelson:2007um} the constraint equation, we only consider
the physical solutions.  The large scale behaviour of the
wave-functions must be normalisable, which is a necessary condition for
having physical wave-functions, while the wave-functions should preserve
{\sl pre-classicality} at large scales, which is a necessary condition for
the validity of the continuum limit. We thus
obtain~\cite{Nelson:2007um} constraints to the two-dimensional
parameter space $(A,\delta)$, shown in Fig.~1 for $A$ in the range
$0<A<-1/2$, imposed from full LQG theory
considerations~\cite{Bojowald:2007ra}.
\begin{figure}
 \begin{center}
  \input{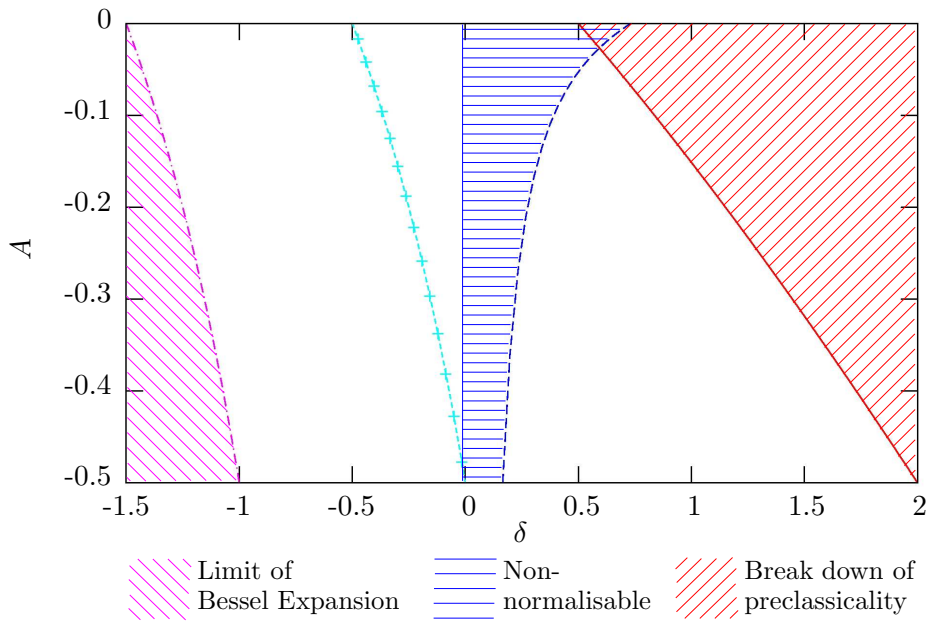}
  \caption{\label{fig6} The allowed types of matter content are
    significantly restricted. For a varying lattice ($A \neq 0$) it is
    not always possible to treat the large scale behaviour of the
    wave-functions perturbatively (dashed line with
    crosses)~\cite{Nelson:2007um}.}
 \end{center}
\end{figure}

In conclusion, the continuum limit of the Hamiltonian constraint
equation is sensitive to the choice of the lattice refinement model
and only a limited range of matter components can be supported within
a particular choice.

\subsection{Uniqueness of the factor ordering in the Wheeler-de Witt equation}

I will now show that the lattice refinement model
$\tilde\mu=\mu_0\mu^{-1/2}$, argued~\cite{ns4} to be the only one
achieved by both physical considerations of large scale physics and
consistency of the quantisation structure, is also the only model
which makes the factor ordering ambiguities of LQC to disappear in the
continuum limit~\cite{ns4}.

Let me be more specific. Writing the gravitational part of the
Hamiltonian constraint in terms of the triad and the holonomies of the
connection, one realises that there are many ways of doing so.
Considering for example,
\begin{equation}\label{eq:ham}
\hat{\mathcal C}_{\rm g} = \frac{2i \ }{\kappa^2 \hbar \gamma^3
  k^3} {\rm tr} \sum_{\rm ijk} \epsilon^{\rm ijk} \left( \hat{h}_{\rm
  i} \hat{h}_{\rm j} \hat{h}_{\rm i}^{-1} \hat{h}_{\rm
  j}^{-1}\hat{h}_{\rm k} \left[ \hat{h}_{\rm k}^{-1},\hat{V} \right]
\right)~, 
\end{equation}
there are  many possible  choices of factor  ordering that  could have
been made  at this point;  classically, the actions of  the holonomies
commute.  However,  each of these  factor ordering choices leads  to a
different  factor ordering  of  the Wheeler-de  Witt  equation in  the
continuum  limit.   The  action  of  the  factor  ordering  chosen  in
Eq.~(\ref{eq:ham}) leads~\cite{ns4} to
\begin{equation}\label{eq:factor1}
\epsilon_{\rm ijk} {\rm tr}  \left( \hat{h}_{\rm i} \hat{h}_{\rm j}
\hat{h}_{\rm i}^{-1} \hat{h}_{\rm j}^{-1}\hat{h}_{\rm k} \left[ 
\hat{h}_{\rm k}^{-1},\hat{V} \right] \right)  =  
-24\widehat{\rm sn}^2\widehat{\rm cs}^2 \left( \widehat{\rm cs} \hat{V}
\widehat{\rm sn} - \widehat{\rm sn} \hat{V} \widehat{\rm cs}\right)~,
\end{equation}
while other choices have certainly a different action.

Defining $\hat{V}|\nu\rangle = V_{\nu}|\nu\rangle$, with $\hat{V}$ the
volume operator with eigenvalues $V_{\nu}$, the action of the above
factor ordering on a general state $|\Psi \rangle = \sum_\nu
\psi_\nu|\nu\rangle$ in the Hilbert space, reads~\cite{ns4}
\begin{eqnarray}\label{eq:ham1.2}
\epsilon_{\rm ijk} {\rm tr}  \left( \hat{h}_{\rm i} \hat{h}_{\rm j}
\hat{h}_{\rm i}^{-1} \hat{h}_{\rm j}^{-1}\hat{h}_{\rm k} \left[ 
\hat{h}_{\rm k}^{-1},\hat{V} \right] \right)|\Psi\rangle & = & 
\frac{-3i}{4} \sum_\nu \Biggl[ \Bigl( V_{\nu-3k} - V_{\nu-5k} \Bigr) 
\psi_{\nu-4k}
-2\Bigl(V_{\nu+k}- V_{\nu-k}\Bigr)\psi_\nu \Biggr. \nonumber \\
&&+\Biggl. \Bigl( V_{\nu+5k} - V_{\nu+3k} \Bigr) \psi_{\nu+4k} \Biggr] |
\nu\rangle~;
\end{eqnarray}
the action of any other factor ordering choice can be obtained in a
similar manner.  By noting that the volume is given by 
\begin{equation}
V_{\nu}|\nu\rangle \sim [\mu\left(\nu\right)]^{3/2}|\nu\rangle~, 
\end{equation}
where $\mu(\nu)$ is obtained by
\begin{equation}
\nu={k\mu^{1-A}\over \mu_0(1-A)}~,
\end{equation}
we find~\cite{ns4} 
\begin{equation} 
V_{\nu\pm nk} \sim \Bigl[ \left(
  \nu \pm nk\right) \alpha \Bigr]^{3/\left[2(1-A)\right]}~, 
\end{equation} 
where $\alpha = \mu_0\left(1-A\right)/k$.

We then take the continuum limit of these expressions by expanding
$\psi_\nu \approx \psi\left(\nu\right)$ as a Taylor expansion in small
$k/\nu$. For the particular factor ordering chosen above, the large
scale continuum limit of the Hamiltonian constraint reads~\cite{ns4}:
\begin{eqnarray}\label{eq:final} 
\lim_{k/\nu \rightarrow 0} \epsilon_{\rm ijk} {\rm tr} \left(
\hat{h}_{\rm i} \hat{h}_{\rm j} \hat{h}_{\rm i}^{-1} \hat{h}_{\rm
  j}^{-1}\hat{h}_{\rm k} \left[ \hat{h}_{\rm k}^{-1},\hat{V} \right]
\right)|\Psi\rangle \sim \nonumber \\ \frac{-36i}{1-A}
\alpha^{3/\left[2\left(1-A\right)\right]} k^3 \sum_\nu
\nu^{\left(1+2A\right)/\left[2\left( 1-A\right)\right]} \Biggl[
  \frac{{\rm d}^2 \psi}{{\rm d} \nu^2 } + \frac{1+2A}{1-A}
  \frac{1}{\nu} \frac{{\rm d} \psi}{{\rm d}\nu} +
  \frac{\left(1+2A\right) \left(4A-1\right)}{\left(1-A\right)^2}
  \frac{1}{4\nu^2} \psi\left(\nu\right) \Biggr]
|\nu\rangle~. \nonumber \\
\end{eqnarray}
One can easily confirm that one obtains the same continuum limit for
the Wheeler-de Witt equation, only for the choice $A=-1/2$~\cite{ns4},
in which case the Wheeler-de Witt equation reads~\cite{ns4}
\begin{equation}\label{eq:final_con} 
\lim_{k/\nu \rightarrow 0} {\cal C}_{\rm g} |\Psi\rangle =
\frac{72}{\kappa^2 \hbar \gamma^3}
\left(\frac{\kappa\gamma\hbar}{6}\right)^{3/2} \sum_\nu \frac{{\rm
    d}^2 \psi}{{\rm d} \nu^2} |\nu\rangle~.  
\end{equation}
Thus, there is only one lattice refinement model, namely
 $\tilde\mu=\mu_0\mu^{-1/2}$, with a non-ambiguous continuum
 limit.

In conclusion, phenomenological and consistency requirements lead to a
particular lattice refinement model, implying that LQC predicts a
unique factor ordering of the Wheeler-de Witt equation in its
continuum limit.  Alternatively, demanding that factor ordering
ambiguities disappear in LQC at the level of Wheeler-de Witt equation
leads to a unique choice for the lattice refinement model.

\subsection{Numerical techniques in solving the Hamiltonian constraint}

Lattice refinement leads to new dynamical difference equations, which
being of a non-uniform step-size, imply technical complications. More
precisely, the information needed to calculate the wave-function at a
given lattice point is not provided by previous iterations. This
becomes clear in the case of two-dimensional wave-functions, as for
instance in the study of Bianchi models or black hole interiors.  I
will present below a method~\cite{ns3} based on Taylor expansions that
can be used to perform the necessary interpolations with a
well-defined and predictable accuracy. 

For a one-dimensional difference equation defined on a varying
lattice, the Hamiltonian constraint can be mapped onto a fixed lattice
simply by a change of basis~\cite{ns3}. This method is however not
useful for the two-dimensional case, where the Hamiltonian constraint
is a difference equation on a varying lattice~\cite{Bojowald:2007ra}:
\begin{eqnarray}
\label{eq:vary2D}
&&C_{+}\left(\mu,\tau\right) \left[
  \Psi_{\mu+2\delta_\mu,\tau+2\delta_\tau} - \Psi_{\mu-2\delta_\mu,
    \tau+2\delta_\tau} \right] \nonumber \\ 
&&+ C_0\left(\mu,\tau\right) \left[
  \left(\mu+2\delta_\mu\right)\Psi_{\mu+4\delta_\mu, \tau} - 2\left( 1
  + 2 \gamma^2 \delta_\mu^2\right) \mu\Psi_{\mu,\tau}
  +\left(\mu-2\delta_\mu\right)\Psi_{\mu-4\delta_\mu,\tau} \right]
\nonumber \\ 
&&+C_{-} \left( \mu,\tau\right) \left[
  \Psi_{\mu-2\delta_\mu,\tau-2\delta_\tau} -
  \Psi_{\mu+2\delta_\mu,\tau-2\delta_\tau}\right] = \frac{\delta_\tau
  \delta_\mu^2}{\delta^3} {\cal H}_\phi\Psi_{\mu,\tau}~, 
\end{eqnarray}
with
\begin{eqnarray}
C_{\pm} &\equiv& 2\delta_\mu \left( \sqrt{\left| \tau \pm 2
  \delta_\tau \right| } + \sqrt{\left| \tau\right| } \right)~, \\ 
C_0 &\equiv& \sqrt{\left| \tau + \delta_\tau\right|} - \sqrt{ \left| \tau -
  \delta_\tau\right| }~, 
\end{eqnarray} 
where $\delta_\mu$ and $\delta_\tau$ denote the step-sizes along the
$\mu$ and $\tau$ directions, respectively.  

\begin{figure}
 \begin{center}
   \includegraphics{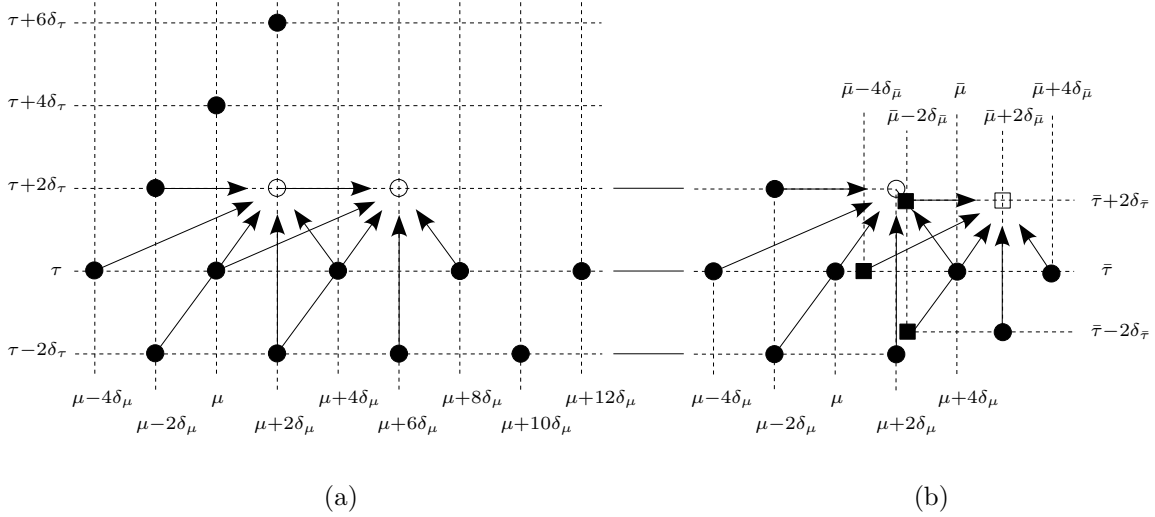}
   \caption{\label{fig5} (a) For the fixed lattice case the
     two-dimensional wave-function can be calculated, given suitable
     initial conditions (solid circles).  (b) In the case of a
     refining lattice, the data needed to calculate the value of the
     wave-function at a particular lattice site (open square) is not
     given by previous iterations (solid squares)~\cite{ns3}.}
 \end{center}
\end{figure}

In the case of lattice refinement, $\delta_\mu$ and $\delta_\tau$ are
decreasing functions of $\mu$ and $\tau$, respectively, and the data
needed to calculate the value of the wave-function at a particular
lattice site are not given by previous iterations, as it is
illustrated in Fig.\ref{fig5}.
We propose~\cite{ns3} to use Taylor expansions to calculate the
necessary data points.  Let us assume that the matter Hamiltonian acts
diagonally on the basis states of the wave-function, namely
\begin{equation}
\hat{\cal H}_\phi |\Psi \rangle \equiv \hat{\cal H}_\phi
\sum_{\mu,\tau} \Psi_{\mu,\tau} |\mu,\tau\rangle = \sum_{\mu,\tau}
    {\cal H}_\phi \Psi_{\mu,\tau} |\mu,\tau \rangle~.  
\end{equation}
Given a function evaluated at three (non-collinear) coordinates, the
Taylor approximation to the value at a fourth position is 
\begin{eqnarray}
 f\left( x_4,y_4\right) &=& f\left(x_2, y_2\right) + \delta^x_{42}
 \frac{\partial f}{\partial x} \Big|_{x_2, y_2} + \delta^y_{42}
 \frac{\partial f}{\partial y} \Big|_{x_2, y_2} \nonumber\\ 
&& + {\cal O}\left( \left(\delta^x_{42}\right)^2 \frac{\partial^2
   f}{\partial x^2}\Big|_{x_2,y_2}\right)+{\cal O}\left(
 \left(\delta^y_{42}\right)^2 \frac{\partial^2 f}{\partial
   y^2}\Big|_{x_2,y_2}\right) ~,\label{eq:taylor} 
\end{eqnarray}
where the Taylor expansion is taken about the position
$\left(x_2,y_2\right)$, we have defined $\delta_{ij}^x \equiv x_i-x_j$ and
$\delta_{ij}^y \equiv y_i-y_j$, and the differentials can be approximated using
\begin{eqnarray}
f\left(x_1,y_1\right) &=& f\left(x_2,y_2\right) + \delta^x_{12} \frac{
  \partial f}{\partial x}\Big|_{ x_2,y_2} + \delta^y_{12}
\frac{\partial f}{\partial y}\Big|_{x_2,y_2} + \cdots~,\\
f\left(x_3,y_3\right) &=& f\left(x_2,y_2\right) + \delta^x_{32} \frac{
  \partial f}{\partial x}\Big|_{ x_2,y_2} + \delta^y_{32}
\frac{\partial f}{\partial y}\Big|_{x_2,y_2} + \cdots~,
\end{eqnarray}
where the dots indicate higher order terms.

For slowly varying wave-functions, linear approximation is very
accurate and higher order terms in Taylor expansion can only improve
the accuracy by $10^{-2}\%$~\cite{ns3}.  This method can be applied in
any lattice refinement model, while its accuracy can be estimated.

By using this Taylor expansions method, we confirmed~\cite{ns3}
numerically the stability criterion of the Schwarzchild interior,
found~\cite{Bojowald:2007ra} earlier using a von Neumann analysis, and
investigated~\cite{ns3} the way that lattice refinement modifies the
stability properties of the system.

Let me now show how the underlying discreteness of space-time leads to
a twist~\cite{ns3} in the wave-functions, for both a fixed and a
varying lattice model.  In the case of a constant lattice, an
initially centred Gaussian will move to larger $\mu$, as $\tau$ is
increased~\cite{ns3}.  However, for regions where the lattice
discreteness is important, this no longer holds, and the value at one
lattice point introduces a non-zero component to the value at a
lattice point with larger $\mu$ coordinate, {\sl i.e.}, the
wave-function moves to larger $\mu$. This implies the existence of
some induced rotation on the wave-function due to the underlying
discreteness of the space-time.  Including lattice refinement, this
effect persists~\cite{ns3}. In other words, there is no motion for
$\tau \gg \delta_\tau$, while in the case of lattice refinement this
requirement is reached for lower $\tau$, since $\delta_\tau$ reduces
as $\tau$ increases. As the wave-function moves into a region in which
the discreteness of the lattice is important, a motion will be
induced, as in the constant lattice case.

\section{Anisotropic LQC}

Various aspects of anisotropic cosmologies have been studied within
LQC in the past~\cite{Bojowald:2007ra,Chiou:2007sp}, however the first
full and consistent quantisation of a Bianchi~I cosmology (the
simplest of anisotropic cosmological models) was achieved recently in
Ref.~\cite{Ashtekar:2009vc}.  Moreover, the link back to the
underlying full LQG theory has been strengthened, by considering the
flux of the triads through surfaces consistent with the Bianchi~I
anisotropic case.  With the quantisation of the Bianchi~I model under
control, it is possible to ask whether LQC features, obtained within
the context of isotropic Friedmann-Lema\^{i}tre-Roberston-Walker
cosmology, are robust, at least with respect to this limited extension
of the symmetries of the system.  Bianchi type~I models, apart their
simplicity, they present a particular interest for space-like
singularities within the full LQG theory. Following the same vein as
for the isotropic case, a massless scalar field plays the r\^ole of an
internal time parameter. In the absence of such a field, physical
evolution can be found by constructing families of unitarily related
partial observables, parametrised by geometry degrees of
freedom~\cite{mgt}.

\subsection{Unstable Bianchi~I LQC}

Studying stability conditions of the full Hamiltonian constraint
equation describing the quantum dynamics of the diagonal Bianchi I
model, I will show~\cite{inst} that there is robust evidence of an
instability in the explicit implementation of the difference equation.
On the one hand, such a result may question the choice of the
quantisation approach, the model of lattice refinement, and/or the
r\^ole of ambiguity parameters. On the other hand, one may argue that
such an instability may not be necessarily a problem since it might be
that unstable trajectories are explicitly removed by the physical
inner product.

Consider the difference equation arising from the loop quantisation of
the Bianchi~I model~\cite{Ashtekar:2009vc}:
\begin{eqnarray}\label{eq:diff_eqn}
 \partial_T^2 \Psi \left( \lambda_1, \lambda_2, \nu ; T \right) &=&
 \frac{\pi G}{2} \sqrt{\nu} \Bigl[ \left( \nu+2\right) \sqrt{\nu +4}
   \Psi^+_4 \left(\lambda_1,\lambda_2,\nu;T\right) -\left(
   \nu+2\right) \sqrt{\nu} \Psi^+_0\left(
   \lambda_1,\lambda_2,\nu;T\right) \nonumber \\ && - \left(
   \nu-2\right) \sqrt{\nu} \Psi^-_0 \left(
   \lambda_1,\lambda_2,\nu;T\right)
   +\left(\nu-2\right)\sqrt{|\nu-4|}\Psi^-_4 \left(
   \lambda_1,\lambda_2,\nu;T\right) \Bigr]~,
\end{eqnarray}
where
\begin{eqnarray}
 \Psi^+_4 \left(\lambda_1,\lambda_2,\nu;T\right) &=& \sum_{i\neq
   j=(0,1,2)} \Psi \left( a_i \lambda_1, a_j
 \lambda_2,\nu+4;T\right) \nonumber \\
 \Psi^-_4 \left(\lambda_1,\lambda_2,\nu;T\right) &=& \sum_{i\neq
   j=(-3,-2,0)} \Psi \left( a_i \lambda_1, a_j
 \lambda_2,\nu-4;T\right) \nonumber \\
 \Psi^+_0 \left(\lambda_1,\lambda_2,\nu;T\right) &=& \sum_{i\neq
   j=(-1,0,1)} \Psi \left( a_i \lambda_1, a_j \lambda_2,\nu;T\right)
 \nonumber \\
 \Psi^-_0 \left(\lambda_1,\lambda_2,\nu;T\right) &=& \sum_{i\neq
   j=(-2,0,3)} \Psi \left( a_i \lambda_1, a_j \lambda_2,\nu;T\right)~,
\end{eqnarray}
\begin{eqnarray}\label{eq:a}
a_{-3} \equiv \left( \frac{\nu-4}{\nu-2}\right)~,\ \ \ a_{-2} \equiv
\left( \frac{\nu-2}{\nu}\right)~, \ \ \ a_{-1} \equiv \left(
\frac{\nu}{\nu+2}\right)~, \nonumber \\ a_{0} \equiv 1~, \ \ \ a_{1}
\equiv \left( \frac{\nu+4}{\nu+2}\right)~, \ \ \ a_{2} \equiv \left(
\frac{\nu+2}{\nu}\right)~, \ \ \ a_{3} \equiv \left(
\frac{\nu}{\nu-2}\right)~.  
\end{eqnarray}
Note that $\lambda_1, \lambda_2, \lambda_3$ are related to the volume
of the fiducial cell ${\cal V}$ by
\begin{equation}
\hat V\Psi(\lambda_1,\lambda_2,\nu)=2\pi|\gamma|\sqrt{\Delta}|\nu|l^3_{\rm pl}
\Psi(\lambda_1,\lambda_2,\nu)~,
\end{equation}
with $\gamma={\rm sgn}(p_1p_2p_3)|\gamma|$ and 
\begin{equation}
\nu=2\lambda_1\lambda_2\lambda_3~,
\end{equation}

\begin{figure}
 \begin{center}
  \includegraphics[scale=.7]{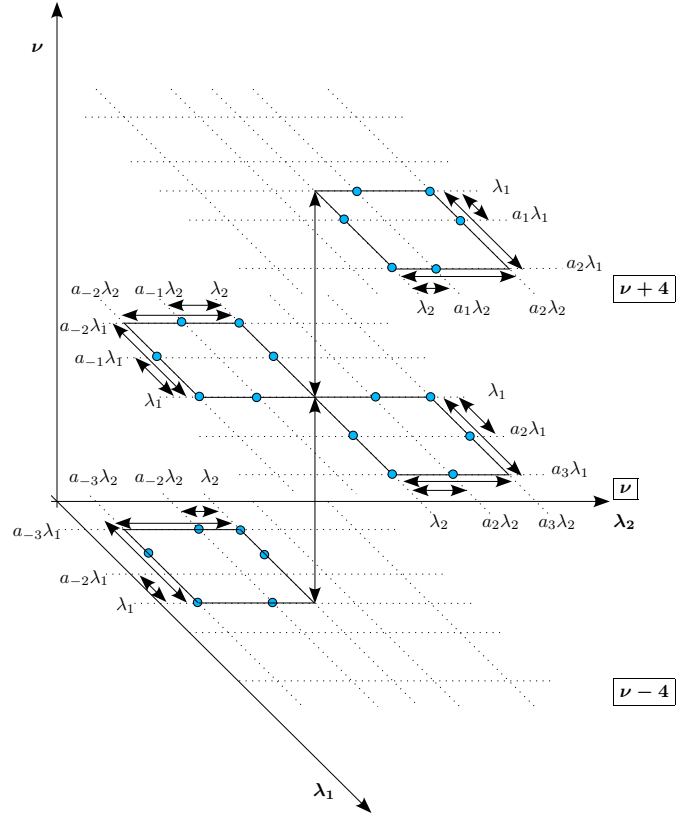}
  \caption{\label{fig1} The geometry of the points used in the
    difference equation that results from the Hamiltonian constraint,
    for the Bianchi~I model~\cite{inst}.}
 \end{center}
\end{figure}

We will investigate the stability of the vacuum solutions, in which
case the solution is static, namely
$\Psi\left(\lambda_1,\lambda_2,\nu;T\right) =
\Psi\left(\lambda_1,\lambda_2,\nu\right)$, simplifying considerably
Eq.~(\ref{eq:diff_eqn}), whose geometry is drawn in
Fig.~\ref{fig1}. To search for growing mode solutions we will perform
a von Neumann analysis.
In addition to specifying the boundary conditions on the $\nu$ and
$\nu-4$ planes, we are also required to specify the value at five of
the points given in
$\Psi^+_4\left(\lambda_1,\lambda_2,\nu\right)$. There are in total
$23$ values that are required and with such initial data the
difference equation, Eq.~(\ref{eq:diff_eqn}), can be used to evaluate
the $24^{\rm th}$ point. Once this point has been evaluated, it can be
used to {\sl move} the central point and evaluate the wave-function at
subsequent positions in the $\nu+4$ plane.

Following the standard von Neumann stability analysis, we decompose
the solutions of the difference equation into Fourier modes and look
for growing modes. Considering the ansatz~\cite{inst}
\begin{equation}
 \Psi\left(\lambda_1,\lambda_2,\nu\right) = T\left(\lambda_1\right)
 \exp \left( i \left( \omega \lambda_2 + \chi \nu \right)\right)~, 
\end{equation}
where we have chosen the $\lambda_1$ direction to be the direction in
which the $\nu+4$ plane is {\sl evolved}, the difference equation
becomes
\begin{eqnarray}
\label{dif-eq}
 e^{4\chi i} \sum_{i\neq j = \left(0,1,2\right)} T\left(a_i
\lambda_1\right) e^{ i\left( \omega a_j \lambda_2 + \chi \nu\right)}&=&
\sqrt{ \frac{\nu}{\nu + 4} } \sum_{i\neq j =\left( -1,0,1\right)}
T\left( a_i \lambda_1\right) e^{i\left(\omega a_j \lambda_2 + \chi
  \nu\right)} \nonumber \\
&& + \left(\frac{\nu-2}{\nu+2}\right) \sqrt{ \frac{\nu}{\nu + 4} }
 \sum_{i\neq j =\left( -2,0,3\right) } T\left( a_i \lambda_1\right)
 e^{i\left(\omega a_j \lambda_2 + \chi \nu\right)}\nonumber\\
&& -\left(\frac{\nu-2}{\nu+2}\right)\sqrt{ \frac{|\nu-4|}{\nu + 4} }
 e^{-4\chi i} \sum_{i\neq j =\left( -3,-2,0\right) } T\left( a_i
 \lambda_1\right) e^{i\left(\omega a_j \lambda_2 + \chi \nu\right)}~.
 \nonumber \\
\end{eqnarray}
As it has been explicitly shown in Ref.~\cite{inst}, expanding in
terms of small $1/\nu$, the difference equation can be written in the
form of a vector equation as
\begin{equation} 
M_1 \overline{T}_{3} = M_2 \overline{T}_{2}~, 
\end{equation} 
where we have defined the vectors 
\begin{equation} 
\overline{T}_i = \left[ \begin{array}{c} T\left(a_i \lambda_1\right)
    \\ T\left(a_{i-1} \lambda_1\right) \\ T\left(a_{i-2}
    \lambda_1\right) \\ T\left(a_{i-3} \lambda_1\right)
    \\ T\left(a_{i-4} \lambda_1\right) \\ T\left(a_{i-5}
    \lambda_1\right) \end{array} \right]~~\mbox{for}~~i=2,3 
\end{equation} 
and the matrices 
\begin{equation} 
M_1 = \left( \begin{array}{cccccc} A & 0 & 0 & 0 & 0 & 0 \\ 0 & 1 & 0
  & 0 & 0 & 0 \\ 0 & 0 & 1 & 0 & 0 & 0 \\ 0 & 0 & 0 & 1 & 0 & 0 \\ 0 &
  0 & 0 & 0 & 1 & 0 \\ 0 & 0 & 0 & 0 & 0 & 1 \end{array} \right)~,
\ \ \ \ \
M_2 = \left( \begin{array}{cccccc}
B & C & D & E & F & G \\
1 & 0 & 0 & 0 & 0 & 0 \\
0 & 1 & 0 & 0 & 0 & 0 \\
0 & 0 & 1 & 0 & 0 & 0 \\
0 & 0 & 0 & 1 & 0 & 0 \\
0 & 0 & 0 & 0 & 1 & 0 \end{array} \right)~.
\end{equation}
The condition for stability can be written as follows: If
\begin{equation}
 \max |\tilde{\lambda} | \leq 1 ~~~\forall~~\omega\ \mbox{and}\ \chi~,
\end{equation}
where $\tilde{\lambda}$ are the eigenvalues of the matrix
$\left(M_1\right)^{-1}M_2$, the amplitude $T\left(a_3\lambda_1\right)$
is less than that of previous points, in other words, the difference
equation is stable. Let us define the parameter
$\Lambda=\omega\lambda_2/\nu$.  The presence of an instability in the
difference equation has been demonstrated~\cite{inst} in several ways:
(i) there is a particular set of critical modes,
$\Lambda=(2n-1)\pi/2$, with $n\in \mathbb{Z}$, for which the system is
unstable; (ii) in the large $\nu$ limit, the system is unstable for
the modes $\Lambda = \pi/4$ and $\chi = 0$; (iii) the system is
unstable for a general $\nu$, for modes that approach the critical
value.

In conclusion, the difference equation, Eq.~(\ref{eq:diff_eqn}), is
unconditionally unstable, in the sense that there is no region of
$\left(\lambda_1,\lambda_2,\nu\right)$ in which
Eq.~(\ref{eq:diff_eqn}) is stable.

\subsection{Lattice refinement from isotropic embedding of anisotropic 
cosmology}

Given a consistent quantum anisotropic model, one can
find~\cite{iso-emb} isotropic states for which the discrete step-size
of the isotropically embedded Hamiltonian constraint is not
necessarily that of the $p^{-1/2}$ ({\sl new quantisation}) step-size.
The choice of different embeddings has important consequences for the
precise form of discretisation in the isotropic sub-system. In this
sense, lattice refinement could be interpreted as being due to the
degrees of freedom that are absent in the isotropic
model~\cite{iso-emb}.

In the standard approach, isotropic states are taken to be those in
which the three scale factors are $\lambda_1 = \lambda_2 = \lambda_3$,
or more precisely, defining the volume of the state as $\nu = 2
\lambda_1\lambda_2\lambda_3$ to eliminate one of the directions
($\lambda_3$), the map
\begin{equation} 
|\Psi \left( \lambda_1, \lambda_2, \nu\right)\rangle \rightarrow
\Big|\sum_{\lambda_1,\lambda_2} \Psi \left( \lambda_1, \lambda_2,
\nu\right)\rangle \equiv |\Psi \left(\nu\right)\rangle~, 
\end{equation} 
produces isotropic states. Working with the three scale factors
$\lambda_i$, one can show~\cite{iso-emb} that there is an ambiguity in
exactly what the volume of such an isotropic state is.
Consider the state
\begin{equation}
\label{eq:state1} 
|\tilde\Psi \left( \lambda_1,\lambda_2,\lambda_3\right)
\rangle \equiv \frac{1}{A}\Bigl[ |\Psi \left(
  \lambda_1,\lambda_2,\lambda_3\right)\rangle + |\Psi \left(
  \lambda_3,\lambda_1,\lambda_2\right)\rangle + |\Psi \left(
  \lambda_2,\lambda_3,\lambda_1\right)\rangle \Bigr]~, 
\end{equation}
with the restriction
\begin{equation}
 \langle \Psi\left(\lambda_1,\lambda_2,\lambda_3\right) 
| \Psi
\left(\lambda_3,\lambda_1,\lambda_2\right)\rangle  = \langle
\Psi\left(\lambda_1,\lambda_2,\lambda_3\right) | \Psi
\left(\lambda_2,\lambda_3,\lambda_1\right)\rangle \nonumber  =
\langle \Psi\left(\lambda_3,\lambda_1,\lambda_2\right) | \Psi
\left(\lambda_2,\lambda_3,\lambda_1\right)\rangle\nonumber~, 
\end{equation}
on the anisotropic states.  The expectation values of the scale
factors along each direction of such a state are 
\begin{equation} 
\langle \hat{\lambda}_i \rangle = \frac{\lambda_1 + \lambda_2 +
  \lambda_3}{3}~.  
\end{equation} 
The measured scale factor is equal in each direction and is given by
the average of the scale factors of the underlying, anisotropic
states. However, the measured volume of such a state is just $\nu =
2\lambda_1\lambda_2\lambda_3$, which is {\it not} the cube of the
measured scale factor. Thus, while it is the eigenvalue of the
anisotropically defined volume operator, it is not necessarily what we
would measure as the volume. Essentially, the reason for this is that
while the scale factors $\lambda_i$ are measured to be equal in each
direction, they are not eigenvalues of the state, {\it i.e.},
$\hat{\lambda}_i
|\tilde\Psi\left(\lambda_1,\lambda_2,\lambda_3\right)\rangle \neq
\lambda_i|\tilde\Psi\left(\lambda_1,\lambda_2,\lambda_3\right)\rangle$. However,
both the average and the product ({\sl i.e.}, the volume) of the scale
factors are eigenvalues. It is this ambiguity, that leads to the
possibility of deviations from the standard isotropic case. 

In conclusion, the difference between the two procedures is
essentially due to what one considers to be more fundamental, the
volume of the underlying states ($\nu$) or the measured volume of the
symmetric state ($\langle \lambda \rangle ^3$), which are not
necessarily equal. Choosing $\nu$ leads to the {\sl new} quantised
Hamiltonian of isotropic cosmology, while choosing $\langle \lambda
\rangle ^3$ results in some kind of different {\sl lattice
refinement}. This {\sl lattice refinement} is significantly more
complicated that the single power law behaviour, $p^{{\cal A}}$,
usually considered.

\section{Conclusions}

Loop Quantum Gravity proposes a method of quantising gravity in a
background independent, non-perturbative way. Quantum gravity is
essential when curvature becomes large, as for example in the early
stages of the evolution of the universe.  Applying LQG in a
cosmological context leads to Loop Quantum Cosmology which is a
symmetry reduction of the infinite dimensional phase space of the full
theory, allowing us to study certain aspects of the theory
analytically.  The discreteness of spatial geometry, a key element of
the full theory, leads to successes in LQC which do not hold in the
Wheeler-de Witt quantum cosmology. 

By studying phenomenological consequences of LQC we can get some
useful insight about the full LQG theory and get an answer to some
long-standing questions in early universe cosmology.  Here, I have
briefly described some of the phenomenological aspects of LQC.

\ack It is a pleasure to thank the organisers of the First
Mediterranean Conference on Classical and Quantum Gravity (MCCQG)
(http://www.phy.olemiss.edu/mccqg/), for inviting me to give
this talk, in the beautiful island of Crete  in Greece.  This work
is partially supported by the European Union through the Marie Curie
Research and Training Network {\sl UniverseNet} (MRTN-CT-2006-035863).


\section*{References}
\begin{thebibliography}{9}
\bibitem{infl} Guth A 1981 {\it Phys. Rev. D} {\bf 23} 347
\bibitem{drawbacks} Sakellariadou M 2008 {\it Lect. Notes Phys.} {\bf
  738} 359
\bibitem{generic1} Piran T 1986 {\it Phys. Lett. B} {\bf 181} 238
\bibitem{generic2} Goldwirth D 1991 {\it Phys. Rev. D} {\bf 43} 3204
\bibitem{generic3} Calzetta E and Sakellariadou M 1992 {\it
  Phys. Rev. D} {\bf 45} 2802
\bibitem{generic4} Calzetta E and Sakellariadou M 1993 {\it
  Phys. Rev. D} {\bf 47} 3184
\bibitem{generic5} Germani C, Nelson W and Sakellariadou M 2007 {\it
  Phys. Rev. D} {\bf 76} 043529
\bibitem{generic6} Gibbons G and Turok N 2008 {\it Phys. Rev. D}
  {\bf 77} 063516
\bibitem{generic7} Ashtekar A and Sloan D 2009 Loop quantum cosmology
  and slow roll inflation ({\sl Preprint} 0912.4093)
\bibitem{rovelli04} Rovelli C 2004 {\it Quantum Gravity} (Cambridge:
  Cambridge University Press)
\bibitem{thiemann07} Thiemann T 2007 {\it Modern Canonical Quantum
  General Relativity} (Cambridge: Cambridge University Press)
\bibitem{bojowald05} Bojowald M 2005 {\it Living Rev. Rel.} {\bf 8}
  11
\bibitem{al} Ashtekar A and Lewandowski J 1997 {\it Class.\ \&
  Quant.\ Grav.\ } {\bf 14} A55
\bibitem{Ashtekar:2006wn} Ashtekar A, Pawlowski T and Singh P 2006
  Phys.\ Rev.\ D {\bf 74} 084003
\bibitem{Vandersloot:2005kh} Vandersloot K 2005 {\it Phys.\ Rev.\ D}
  {\bf 71} 103506
\bibitem{Rosen:2006bga} Rosen J, Jung J H and Khanna G 2006
  {\it Class.\ Quant.\ Grav.} {\bf 23} 7075 
\bibitem{Bojowald:2007ra}
  Bojowald M, Cartin D and Khanna G 2007
  Phys.\ Rev.\  D {\bf 76} 064018
\bibitem{ms-review}
Sakellariadou M 2009 {\it J.\ Phys.\ Conf.\ Ser.} {\bf 189} 012035
\bibitem{Nelson:2007um} Nelson W and Sakellariadou M (2007) {\it
  Phys.\ Rev.\ D} {\bf 76} 104003
\bibitem{ns2} Nelson W and Sakellariadou M 2007 {\it Phys.\ Rev.\ D}
  {\bf 76} 044015
\bibitem{gs08} Corichi A and Singh P 2008 {\it Phys.\ Rev.\ D} {\bf
  78} 024034
\bibitem{ns4} Nelson W and Sakellariadou M 2008 {\it Phys.\ Rev.\ D}
  {\bf 78} 024006
\bibitem{Ashtekar:2006uz2} Ashtekar A, Pawlowski T and Singh P 2006
  {\it Phys.\ Rev.\ D} {\bf 74}, 084003
\bibitem{ns3} Nelson W and Sakellariadou M 2008 {\it Phys.\ Rev.\ D}
  {\bf 78} 024030
\bibitem{Chiou:2007sp} Chiou D W and Vandersloot K 2007
{\it Phys.\ Rev.\ D} {\bf 76} 084015
\bibitem{Ashtekar:2009vc} Ashtekar A and Wilson-Ewing E 2009
 {\it Phys.\ Rev.\ D} {\bf 79} 083535
\bibitem{mgt}
Martin-Benito M, Mena Marugan G A and Pawlowski T 2009
{\it Phys.\ Rev.\ D} {\bf 80} 084038
\bibitem{inst} Nelson W and Sakellariadou M 2009 {\it
  Phys.\ Rev.\ D} {\bf 80} 063521
\bibitem{iso-emb} Nelson W and Sakellariadou M 2009  Lattice
  Refining Loop Quantum Cosmology from an Isotropic Embedding of
  Anisotropic Cosmology ({\sl Preprint} 0906.0292)

\end{thebibliography}
\end{document}